\documentclass[%
 aip,
amsmath,amssymb,
reprint,
]{revtex4-1}
\usepackage{graphicx}
\usepackage{dcolumn}
\usepackage{bm}
\usepackage{siunitx}
\usepackage[utf8]{inputenc}
\usepackage[T1]{fontenc}
\usepackage{mathptmx}
\usepackage{etoolbox}
\usepackage{overpic}
\usepackage{pict2e}
\usepackage[wby]{callouts}
\usepackage{lineno}

\makeatletter
\def\@email#1#2{%
 \endgroup
 \patchcmd{\titleblock@produce}
  {\frontmatter@RRAPformat}
  {\frontmatter@RRAPformat{\produce@RRAP{*#1\href{mailto:#2}{#2}}}\frontmatter@RRAPformat}
  {}{}
}
\makeatother
\begin{document}

\preprint{AIP/123-QED}
\title[]{Flux coupled tunable superconducting resonator}
\author{Juliang Li}
\email{juliang.li@anl.gov}
\affiliation{Argonne National Laboratory, 9700 South Cass Ave., Lemont, IL, 60439, USA}
\author{Pete Barry}
\affiliation{ Cardiff University, Cardiff CF10 3AT, UK}
\author{Tom Cecil}
\affiliation{Argonne National Laboratory, 9700 South Cass Ave., Lemont, IL, 60439, USA}
\author{Marharyta~Lisovenko}
\affiliation{Argonne National Laboratory, 9700 South Cass Ave., Lemont, IL, 60439, USA}
\author{Volodymyr~Yefremenko}
\affiliation{Argonne National Laboratory, 9700 South Cass Ave., Lemont, IL, 60439, USA}
\author{Gensheng~Wang}
\affiliation{Argonne National Laboratory, 9700 South Cass Ave., Lemont, IL, 60439, USA}
\author{Serhii~Kruhlov}
\affiliation{\parbox[t]{0.8\textwidth}{Department of Physics, Drexel Univeristy,  3141 Chestnut St., Philadelphia, PA 19104, USA}}
\author{Goran~Karapetrov}
\affiliation{\parbox[t]{0.8\textwidth}{Department of Physics, Drexel Univeristy,  3141 Chestnut St., Philadelphia, PA 19104, USA}}

\author{Clarence Chang}
\affiliation{Argonne National Laboratory, 9700 South Cass Ave., Lemont, IL, 60439, USA}
\affiliation{University of Chicago, 5640 South Ellis Ave., Chicago, IL, 60637, USA}
\affiliation{\parbox[t]{0.8\textwidth}{Kavli Institute for Cosmological Physics, U. Chicago, 5640 South Ellis Ave., Chicago, IL, 60637, USA}}
           
\begin{abstract}
We present a design and implementation of frequency-tunable superconducting resonator. The resonance frequency tunability is achieved by flux-coupling a superconducting LC-loop to a current-biased feedline; the resulting screening current leads to a change of the kinetic inductance and shift in the resonance frequency. The thin film aluminum resonator consists of an interdigitated capacitor and thin line inductors forming a closed superconducting loop. The magnetic flux from the nearby current feedline induces Meissner shielding currents in the resonator loop leading to change in the kinetic part of the total inductance of the resonator. We demonstarte continuous frequency tuning within 160 MHz around the resonant frequency of 2.7 GHz. We show that: (1) frequency upconversion is achieved when kHz AC modulation signal is superimposed onto the DC bias resulting in sidebands to the resonator tone; (2) three-wave mixing is attained by parametrically pumping the nonlinear kinetic inductance using a strong RF pump signal in the feedline. The simple architecture is amenable to large array multiplexing and on-chip integration with other circuit components. The concept could be applied in flux magnetometers, upconverters, and parametric ampliﬁers operating above \SI{4}{Kelvin} cryogenic temperatures when alternative high critical temperature material with high kinetic inductance is used.
\end{abstract}
\maketitle

\section{Introduction}
With their high quality factors and easy planar fabrication, superconducting resonators have found wide use in applications such as astronomical detectors\cite{day2003} 
and superconducting quantum computing\cite{wallraff2004, blais2004pra}. Tunable superconducting resonators are attractive as they provide an effective platform for interfacing with various forms of excitation including phonons\cite{teufel2011,han2016}, photons\cite{bochmann2013, balram2016, barzanjeh, andrews2014} and electrical charges\cite{brock2021} where fine frequency tuning is required to achieve desired coupling strength. They also have been developed for characterizing high kinetic inductance superconducting thin films\cite{das2022}.

Common approaches for realizing frequency-tunable microwave resonators include the use of Josephson junctions (JJ)\cite{sandberg2008, wangapl2013, abdo2013, castellanos2007, osborn2007, palacios2008} and high kinetic-inductance superconducting wires\cite{xu2019apl, healey2008, vissers2015, samkharadze2016, kher2016, adamyan2016, luomahaara2014}. The high nonlinear inductance of Josephson junctions offers a large frequency tuning range. However, the junctions have relatively low saturation power due to their limited critical current. On the other hand, high kinetic inductance wires can be easily fabricated because of their simple geometries and are suitable for higher operating temperature and power arising from their higher critical temperature ($T_{c}$) and critical current ($I_{o}$) compared to JJs. Tuning the kinetic inductance requires DC current biasing the inductor, which has typically been realized through direct coupling of the bias current. Here, we present an architecture where this current couples through magnetic flux.

The paper is organized as follows: in Section \ref{sec::design} we discuss the operating principles and design of the resonator along with its fabrication process.  Section \ref{sec::mixing} demonstrates flux coupling of both low frequency ($\sim$10~kHz) and high frequency ($\sim$5 GHz) AC signals which can be used for frequency upconversion and three-wave mixing processes. In section \ref{sec::discus} we describe operating the resonator as parametric amplifier and demonstrating preliminary results showing gain for both degenerate and non-degenerate three-wave pumping. We conclude in Section \ref{sec::con} and discuss potential changes to the resonator to improve the frequency tunability and flux coupling strength.  
\vspace*{-0.3cm}
\section{resonator design}
\label{sec::design}
\begin{figure}
     \begin{overpic}[abs,unit=1pt,scale=.05,width=0.45\textwidth]{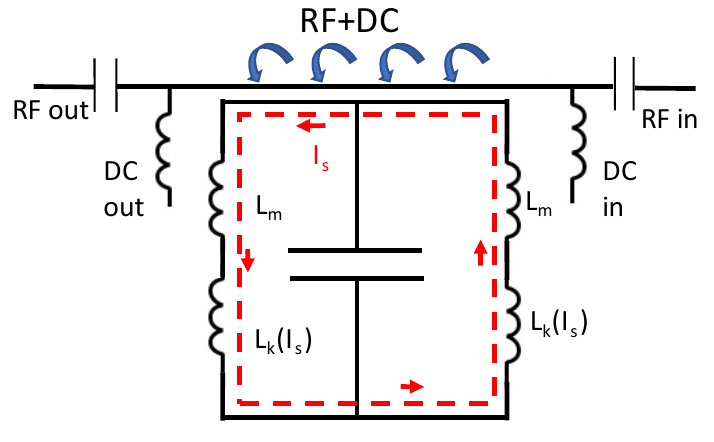}
     \end{overpic}
     \vskip 3pt     
     \caption{\footnotesize{Equivalent circuit of the flux biased tunable resonator. In the center is the interdigitated capacitor and the  inductor carries both geometric ($L_{o}$) and kinetic ($L_{k}$) inductance. RF and DC currents are combined together through bias Tees and fed through the readout line. By Eq.~ \ref{eq::lloopphi}, magnetic flux generated from DC current in the readout line (wide blue arrow bar) excites a screening current $I_{sc}$ (red dash line) in the superconducting loop.}}
\label{fig::design}
\end{figure}

\begin{figure}[h!]
     \begin{overpic}[abs,unit=1pt,scale=.05,angle=0,width=0.45\textwidth]{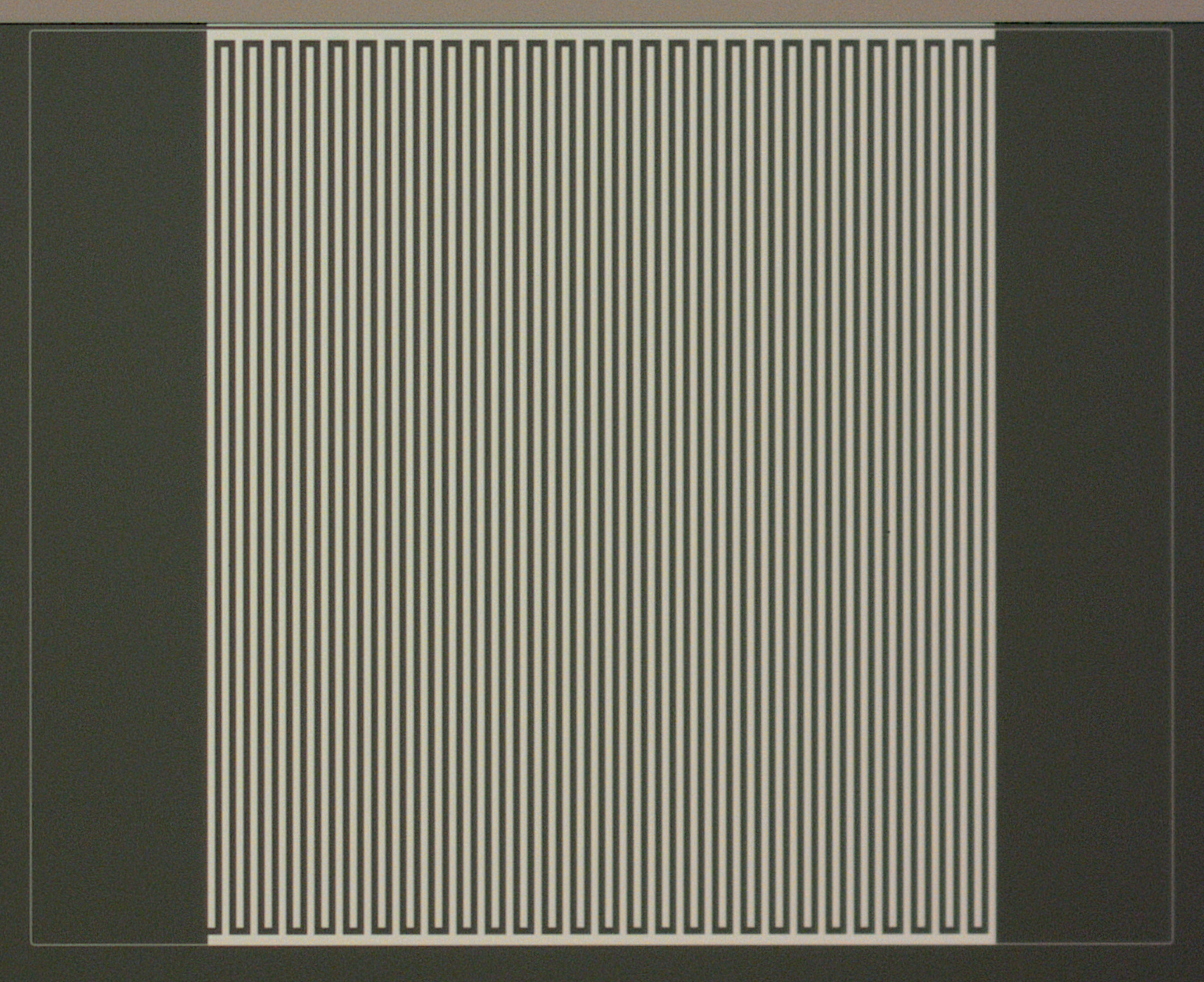}
     \put(0,0){\linethickness{0.35mm}\color{blue}\polygon(2,0)(48,0)(48,28)(2,28)}
     \put(23,167){\linethickness{0.35mm}\color{red}\polygon(0,0)(40,0)(40,20)(0,20)}
     \put(0, -86){\color{blue}\linethickness{0.55mm}
     \frame{\includegraphics[trim={0 0 6cm 0},clip, scale=.059, angle=90]{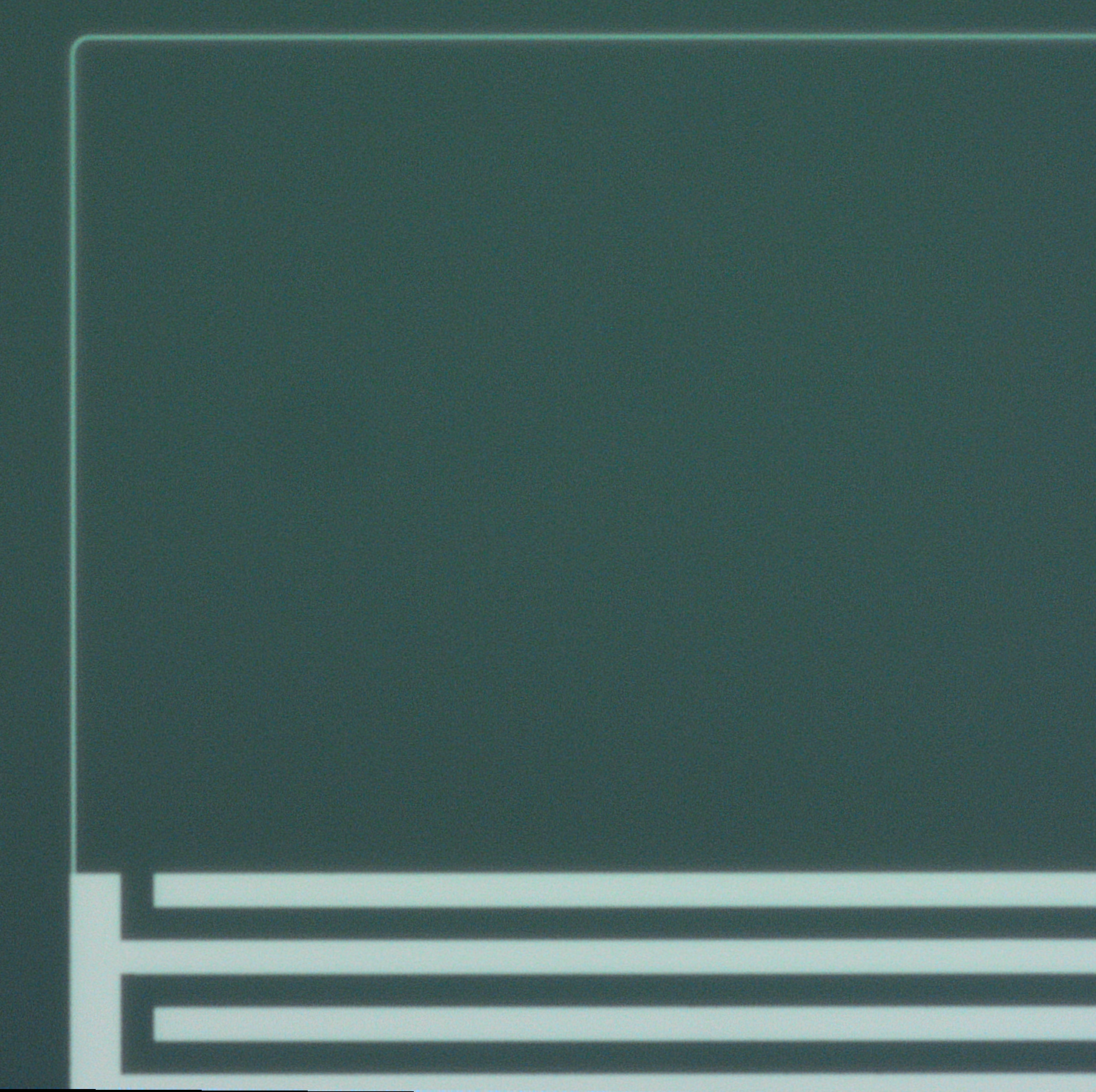}}}
     \put(94, -85){\color{red}\linethickness{0.55mm}
     \frame{\includegraphics[scale=.069, angle=0]{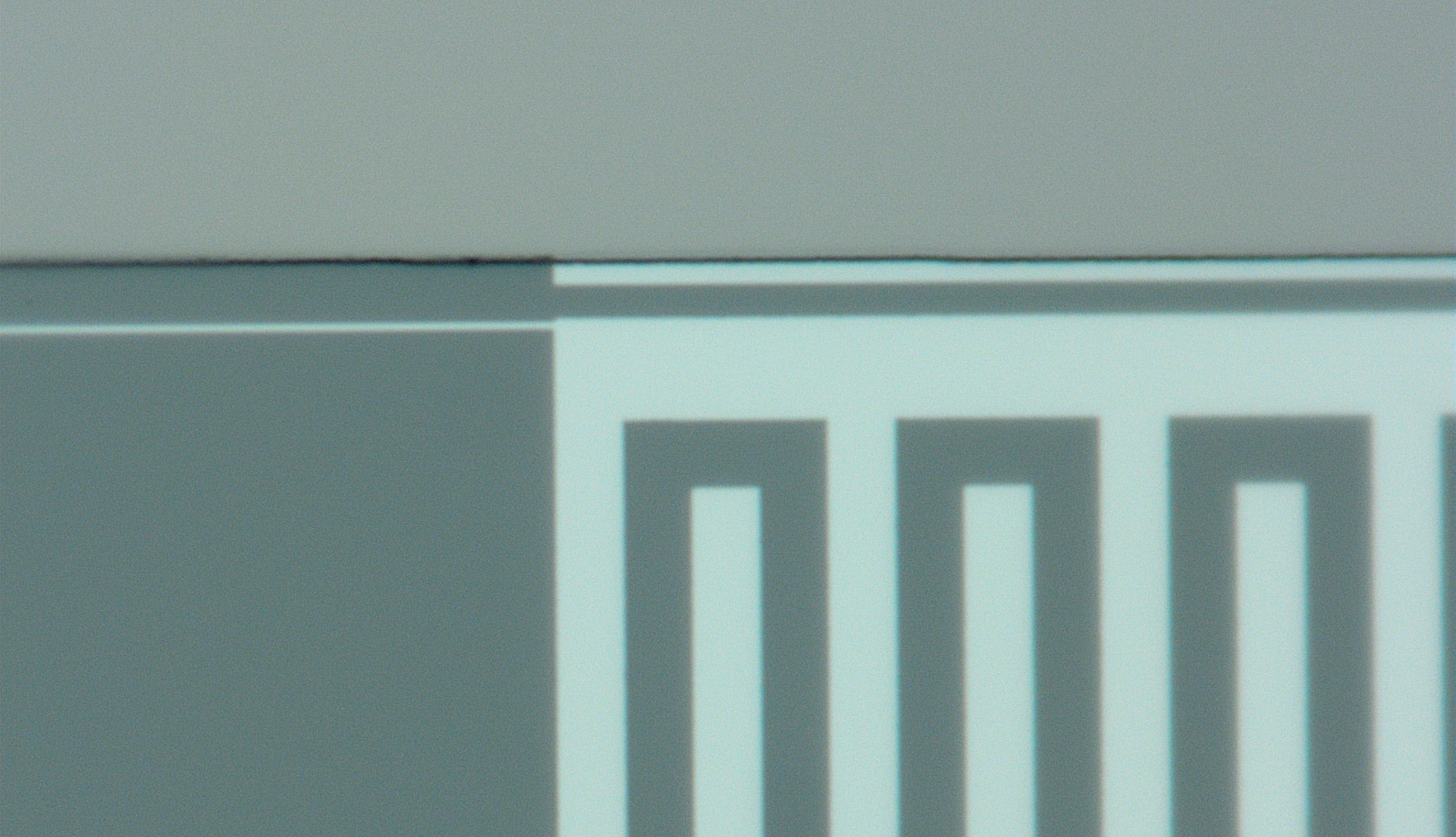}}}
     \put(10,-80){\color{white}\large{(b)}}
     \put(110,-80){\color{white}\large{(c)}}
     \put(110,-25){\color{white}\large{$r_{1}$}}
     \put(10,140){\color{white}\large{(a)}}
     \put(120,-25){\color{white}\vector(1,-0.3){24}}  
     \put(120,-25){\color{white}\vector(1,-0.5){25}}
     \put(200,180){\color{white}\vector(0,-1){80}}  
     \put(200,10){\color{white}\vector(0,1){80}}
     \put(195,93){\color{white}\large{$r_{2}$}}
     \end{overpic}\
     \vskip 105pt
     \caption{\footnotesize{\textbf{a}, image of the tunable resonator. Lighter color marks the Lumped LC resonator and the readout line that carries both the RF and DC signals. Darker color is the silicon substrate. In the center is the interdigitated capacitor and the narrow lines on both sides are the inductors. $r_{2}$ is the distance from the further end of the loop to the readout line that is used for flux calculation. \textbf{b}, zoom in image of the inductor with width of \SI{400}{nm} and interdigitated capacitor in the blue rectangle of \textbf{a}. \textbf{c}, coupling of the LC resonator to the read out line corresponding to the area outlined by the red square of \textbf{a}. Darker color is the Nb readout line and the lighter color is part of the LC resonator. $Q_{c}$ (coupling quality factor) for the resonator is determined by the gap between the resonator and the readout line, which we designed to be \SI{2}{\mu m}. $r_{1}$ is the distance from the closest end of the loop to the readout line that is used for flux calculation.
     }}
\label{fig::chip}
\end{figure}
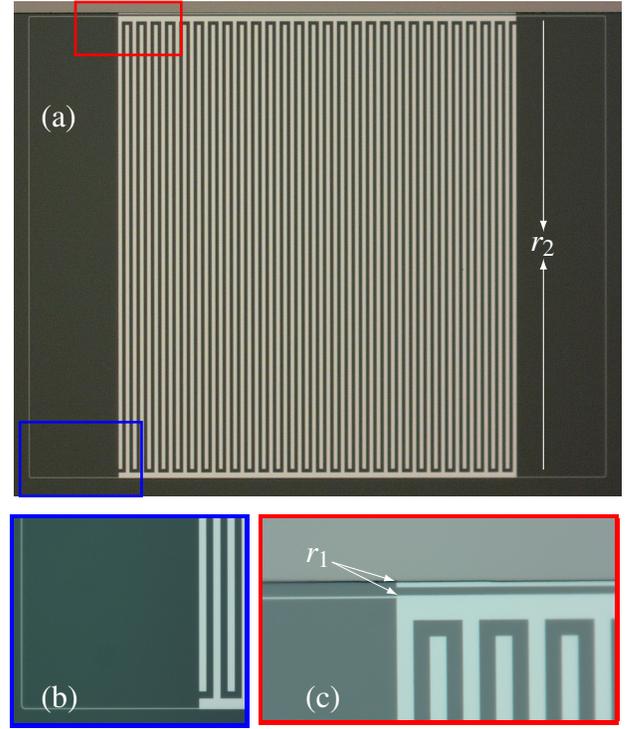

Our tunable resonator design is illustrated in Fig.~\ref{fig::design}. Two inductors are placed in parallel with an interdigitated capacitor in the middle.  The inductors and the two straight edges of the capacitor form a superconducting loop that can carry a continuous circulating Meissner current. Applying current to the readout line via DC `In' and `Out' ports generates a magnetic field and the corresponding total magnetic flux enclosed by the loop given by Eq.~\ref{eq:Philoop}.
\begin{eqnarray}
\Phi_{dc} &=&  w\frac{\mu_{0}I_{dc}}{2\pi}\text{log}\left(\frac{r_2}{r_1}\right).
\label{eq:Philoop}
\end{eqnarray}
where $I_{dc}$ is the DC current supplied through the readout line and $\Phi_{dc}$ is the resulting magnetic flux threading the superconducting loop. $r_{1}$ and $r_{2}$ (shown in Fig.~\ref{fig::chip}) are the distances from the  readout line to the two opposite edges of our square resonator loop that are parallel to the readout line, and $w$ is the width of the square loop parallel to the readout line. 

Flux quantization in the closed superconducting loop neccesitates a circulating, or screening current ($I_{sc}$) given by 
\begin{equation}
    \label{eq::lloopphi}
    L_{\text{self}}\times I_{sc}+\Phi_{\text{dc}} = m\Phi_{o},
\end{equation}
where $L_{\text{self}}$ is the self-inductance of the superconducting loop, $\Phi_{o}$ is the magnetic flux quantum, and $m$ is integer of flux quanta present in the superconducting loop. This screening current will modify the kinetic inductance with relationship given by.
\begin{equation}
L_{k} \approx L_{k_0}\left[1+\left(\frac{I_{sc}}{I^{\ast}}\right)^{2}\right], \label{eq::lkIstar}  
\end{equation}
where $L_{k_0}$ is the intrinsic kinetic inductance at zero current, and $I^{\ast}$ sets the characteristic scale of the nonlinearity and has a value close to the superconducting critical current $I_{c}$\cite{anlage1989, annunziata2010}.

Starting from the design geometry of our resonator, we derive the relative frequency shift as a function of DC bias current  $I_{dc}$ as
\begin{eqnarray}
\frac{\Delta f}{f_{o}}
&=& \frac{\alpha(-M I_{dc}+m\Phi_{0})^{2}}{2L_{\text{self}}^{2}I^{\ast 2}},
\label{eq:freshift}
\end{eqnarray}
where $\alpha$ is kinetic inductance participation ratio, $M$ is the mutual inductance between the readout line and the superconducting loop. Details of the calculation could be found in Appendix~\ref{app::dfdphi}.

Fig.~\ref{fig::chip} shows the physical layout of the frequency tunable microwave resonator. The resonator consists of a central set of interdigitated fingers, which acts as the capacitor, and with both ends connected by narrow lines which act as inductors. The resonator is placed \SI{2}{\mu m} away from a Nb microstrip readout line. The gap provides RF coupling to the feedline which is used for readout of the resonator via a microwave probe tone. Detailed parameters of the design can be found in Table \ref{tab:paras}. Device fabrication is carried out using e-beam lithography and single-layer resist lift-off processes. The process begins with a low resistivity prime silicon wafer with native oxide. The wafer is coated with a 200~nm thick layer of PMMA~950~A4 resist. The resist is exposed in a JEOL JBX-8100FS e-beam writer at a dose of \SI{720}{mJ/cm^{2}} and developed in an MIBK and IPA mix (1:3). Next, a 30~nm thick Al layer is deposited via sputtering at a base pressure below $5\times 10^{-8}$~Torr. The wafer is then placed in Microposit 1165 remover to lift-off the residual Al stack. The readout line is fabricated in a second step with the same lift-off process used on \SI{300}{nm} of Nb. Two resonators are patterned with inductor widths of \SI{400}{nm} and \SI{300}{nm} and the same capacitor dimensions.  

\begin{figure}
\includegraphics[scale=0.6]{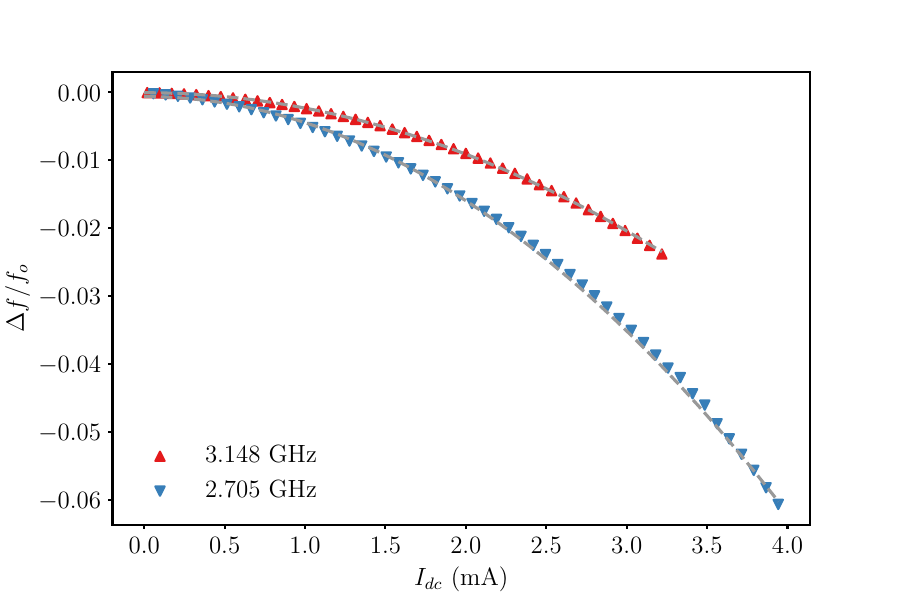}
\put(-200,80){\color{black}\large{$\frac{\Delta f}{f_{o}}\propto \left(\frac{I_{dc}}{I^{\ast}}\right )^{2}$}}
\caption{Relative resonance shift $\Delta f/f_{o}$ as a function of DC bias current for the two Al resonators. The resonant frequency shifts to lower values with increasing biasing current following a parabolic relationship as expected from Eq.~\ref{eq:freshift}. Dashed lines are fits to a quadratic equation.}
\label{fig:foCur}
\end{figure}

\begin{table}
\caption{\label{tab:paras} parameters of the flux biased resonator.}
\begin{ruledtabular}
\begin{tabular}{lc}
 Capacitor: &\\
\hline
finger width ($\mu$m) & 4   \\
finger length ($\mu$m) & 480   \\
gap between fingers ($\mu$m) & 4   \\
gap between finger and capacitor end ($\mu$m) & 4   \\
number of finger pairs & 28 \\
length of capacitor end ($\mu$m) & 500 \\
width of capacitor end ($\mu$m) & 6 \\
\hline
Inductor: & \\
\hline
width (nm) & 300, 400 \\
length ($\mu$m) & 700 \\
\hline
$f_{o}$ (GHz) & 2.705, 3.148 \\
$Q_{o}$ (at df/dI = 0) & 27,000 , 90,000\\
$Q_{c}$ (at df/dI = 0) & 20,000 , 15,000\\
\end{tabular}
\end{ruledtabular}
\end{table}
\label{sec::tuna}
The resonator chip is measured in a dilution refrigerator with base temperature of \SI{30}{mK}. The measured zero-bias resonant frequency for the two resonators, with inductor line width of 400 and \SI{300}~{nm}, are 3.148 and \SI{2.705}~{GHz} respectively.  
Fig.~\ref{fig:foCur} shows the relative frequency shift $\Delta f/f_{o}=(f-f_{o})/f_{o}$ for the two resonators as a function of applied DC current bias. Over \SI{160}{MHz} frequemcy shift is observed for the \SI{2.7}{GHz} resonator. Dashed lines show a fit to a polynomial relationship. Using Eq.~\ref{eq:freshift} we calculate $I^{\ast}$ to be \SI{0.6}{mA} and \SI{0.4}{mA} respectively for resonators with \SI{400}{nm} and \SI{300}{nm} inductors. These values are larger than values estimated through the measured critical current $I_{c}$ (52 and \SI{39}{uA}) and the assumption that $I^{\ast}\approx I_{c}$.

\vspace*{-0.3cm}
\section{mixing through flux biasing}
\label{sec::mixing}
\begin{figure}
         \begin{overpic}[abs,unit=1pt,scale=.15,width=0.5\textwidth]{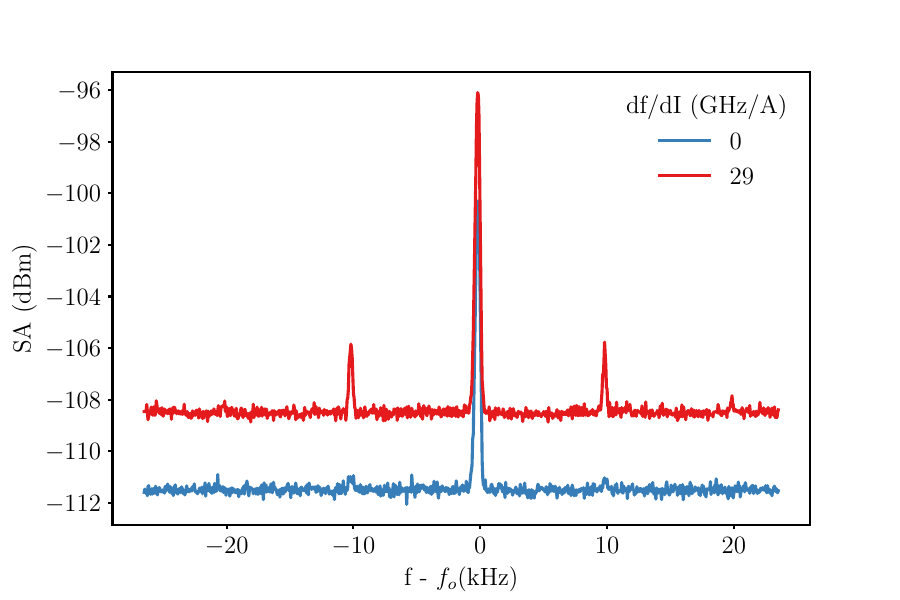}
         \end{overpic}
\caption{\footnotesize{Frequency upconversion of a 10kHz low frequency signal through a carrier tone at resonator frequency $f_{o}$. Without frequency detuning or current bias $df/dI_{dc}=0$ minimum upconversion is observed above the noise floor of our measurement (blue curve). Applying a DC bias that shifts the resonance by 10MHz (corresponding to $df/dI_{dc}\ne 0$) we observe clear frequency upconversion as side bands around the center carrier peak (red curve). }}
\label{fig:modMHz}
\end{figure}

Our flux bias design also presents a mechanism for coupling AC signals. At lower frequencies, this attribute can be used to upconvert the signal. We can consider a low-frequency modulation at frequency $\omega_{m}$ with small amplitude $I_{m}$ superimposed on top of a DC bias for a total current on the readout line $I_{dc}' = I_{dc}+I_{m}\sin(\omega_{m}t)$. Using Eq.~\ref{eq::dfdt} along with the first order Taylor expansion $(I_{dc}+I_{m}\sin(\omega_{m}t))^{2}\approx I_{dc}^{2}+2I_{dc}I_{m}\sin(\omega_{m}t) $ for $I_{m}\ll I_{dc}$, we can write the resonant frequency as:
\begin{eqnarray}
\omega_{o}' &=& \omega_{o} + \delta_{dc}+ \delta_{m},    
\end{eqnarray}
where $\omega_{0}$ is the zero-bias resonant frequency, and 
\begin{eqnarray}
\label{eq::modIdc}
\delta_{dc} &=& -\frac{3\omega_{o}\alpha M^{2}I_{dc}^{2}}{L_{\text{self}}^{2}I_{\ast}^{2}}, \\
\label{eq::modlowf}
\delta_{m} &=& -6\frac{d\omega}{dI_{dc}}I_{m}\sin(\omega_{m}t),
\end{eqnarray}
correspond to the respective shifts in resonantor frequency due to the DC bias and the modulation at $\omega_{m}$. The response to the low frequency modulation, $\delta_{m}$, is a modulation which appears in frequency domain as side bands at $\pm\omega_{m}$ on both sides of the resonator tone $\omega_{o} + \delta_{dc}$ with amplitude proportional to $d\omega/dI_{dc}$. We can demonstrate this upconversion empirically by applying a 10~kHz modulation on the bias line with and without a DC bias. We observe a prominent appearance of $\pm$10~kHz sidebands around the resonator frequency with the application of a DC bias (see Fig.~\ref{fig:modMHz}) while keeping all other measurement parameters the same.
\begin{figure}
         \begin{overpic}[abs,unit=1pt,scale=.095,width=0.52\textwidth]{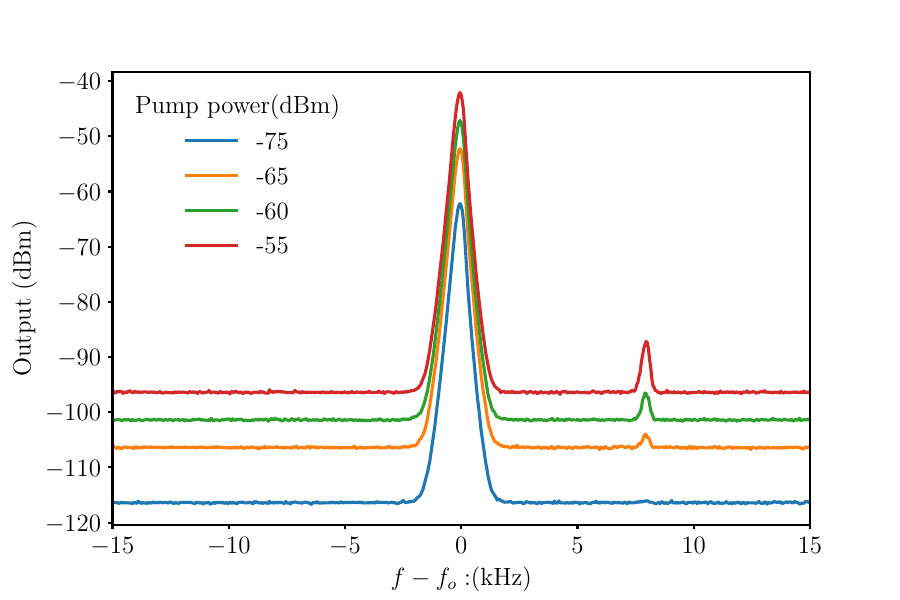}
         \end{overpic}
         \put(-126, 92){\color{white}\linethickness{0.15mm}
     \frame{\includegraphics[trim={0.8cm 0.6cm 0.7cm 0.6cm},clip, scale=.55, angle=0]{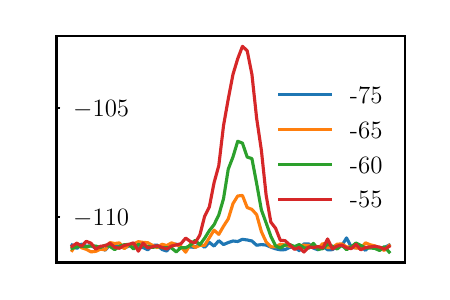}}}
        \caption{\footnotesize{measured three-wave mixing power spectrum. From bottom to top The curves correspond to monotonically increasing pump power. The curves are shifted vertically for better idler tone visibility. Pump power is calculated by subtracting the total lose in the input line from the output power of the signal generator. Both pump and signal generators as well spectrum analyzer are frequency locked through 10MHz reference ports. Inset is zoom in to the idler tone without vertical shift. Horizontal axis is frequency and vertical axis is power in dBm. Legend is pump power in dBm. For pump power from -75 to \SI{-55}{dBm} the idler power increases from close to zero to \SI{10}{dB} relative to the noise floor on the spectrum analyzer.}}
        \label{fig:mixing3and4}
\end{figure}

At higher frequencies, our flux coupling design can be utilized for parametric pumping through three-wave mixing\cite{macklin2015, danie2022}. The system Hamiltonian under three-wave mixing is given by:
\begin{eqnarray}
\mathcal{H} &=& \hbar (\omega_{o}+ \delta_{dc}+\delta_{p}+ K - \frac{\omega_{p}}{2})a^{\dagger} a \nonumber\\ 
&& + \frac{\hbar \xi}{2}a^{\dagger 2} + \frac{\hbar \xi^{\ast}}{2}a^{2} + \frac{\hbar K}{2} a^{\dagger 2} a^{2},
\end{eqnarray}
with 
\begin{eqnarray}
\label{eq:deltaP}
\delta_{p} &=& -\frac{3\omega_{o}\alpha M^{2}I_{p}^{2}}{4L_{\text{self}}^{2}I_{\ast}^{2}}, \\
K &=& -\frac{3\hbar\omega_{o}^{2}\alpha}{2L_{T}I_{\ast}^{2}}, \\
\label{eq::xi}
\xi &=& -\frac{3\omega_{o}\alpha M^{2}I_{dc}I_{p}}{2L_{\text{self}}^{2}I_{\ast}^{2}}e^{-i\phi_{p}}.
\end{eqnarray}
where we have used Eq.~\ref{eq::Iscphi} to convert current in the loop ($\overline{I}_{dc}$) to bias current in the readout line ($I_{dc}$). $\omega_{o}$ is the zero-bias resonance frequency, $L_{T}$ is the total inductance of the inductor which includes both geometric and kinetic inductance, $\delta_{dc}$ is same as in Eq.~\ref{eq::modIdc}, $\delta_{p}$ is the resonance frequency shift from the pumping tone, $K$ is the Kerr constant corresponding to the nonlinearity of the system, $\xi$ is the mixing strength under three-wave mixing and $I_{p}$ is amplitude of the three-wave mixing tone $\omega_{p}$ (see Appendix \ref{ap::resoham} for detailed Hamiltonian construction).

We demonstrate the mixing process as follows: we apply a weak signal tone at frequency $\omega_{s} = \omega_{o} + \delta_{dc}$ and a strong pump tone at frequency $\omega_{p}=2\omega_{s}+\SI{8}{kHz}$ to the readout line and measure the output signal with a spectrum analyzer. Three-wave mixing results in a signal at the idler frequency $\omega_{i}=\omega_{s}+ \SI{8}{kHz}$, which appears as a side band at \SI{8}{kHz} above the signal tone). Fig.~\ref{fig:mixing3and4} shows the results from our measurements with the curves, from bottom up, corresponding to increasing pump power with all other parameters kept constant. The curves have been shifted vertically for clarity. Inset is a zoom in at the idler tone showing that the power at the idler tone increases monotonically with the pump power, which is predicted by the dependence of the three-wave mixing rate $\xi$ on the pump power $I_{p}$ in Eq.~\ref{eq::xi}.
 
\begin{figure}
         \begin{overpic}[abs,unit=1pt,scale=.1,width=0.49\textwidth]{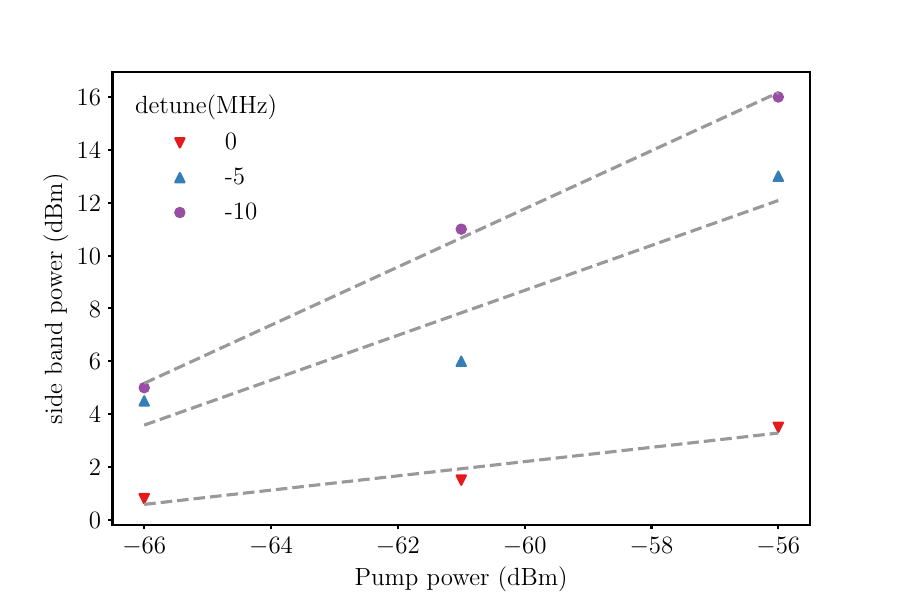}
         \end{overpic} 
        \caption{\footnotesize{Side band power as function of pump power for three different frequency detune or current bias point. In this power measurement we have deducted the noise floor of the spectrum analyzer from  the peak power of the side bands.}}
        \label{fig:mixingslope}
\end{figure}

We repeat this measurement by applying various DC biases which would also change the coupling strength. Results are shown in Fig.~\ref{fig:mixingslope}. From the system Hamiltonian, the strength of the three-wave mixing $\xi$ increases linearly with DC current bias $I_{dc}$ (Eq.~\ref{eq::xi}). With each applied DC bias, we observe an expected shift in resonance frequency given by Eq.~ \ref{eq:deltaP}. We measured the idler sideband power for pump powers of -56, -51 and -46~dBm at three DC biases corresponding to frequency detunings of (0, -5 and \SI{-10}{MHz}). At each detuning we linearly fit the sideband power to its pump power and observe that the slope of the fitted lines increases monotonically with detuning magnitude, which is consistent with the predicted increase in coupling strength.
\vspace*{-0.3cm}
\section{discussion}
\label{sec::discus}
\begin{figure}
         \begin{overpic}[abs,unit=1pt,scale=.52,width=0.49\textwidth]{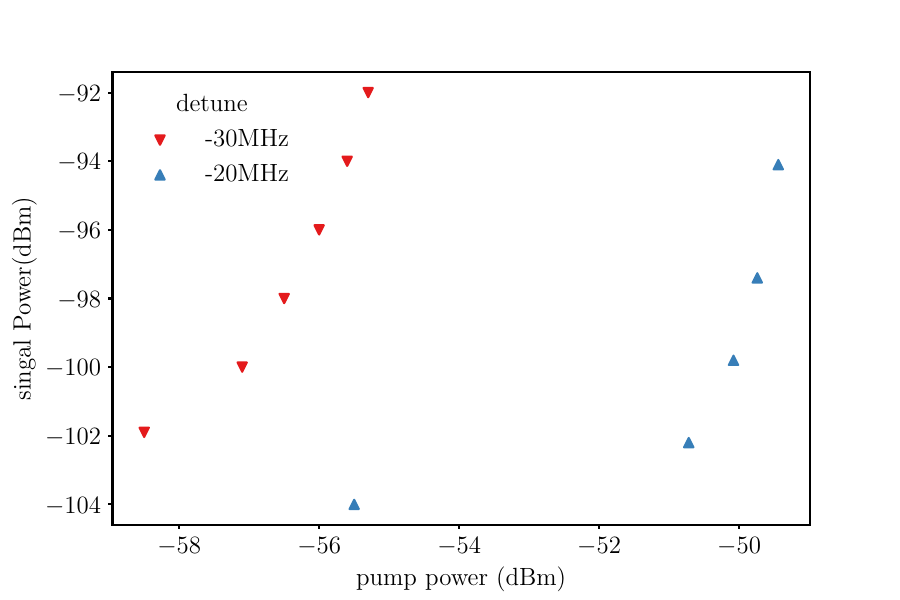}
         \end{overpic}    
        \caption{\footnotesize{Measured signal peak power under non-degenerate three-wave pumping as function of pump power at two frequency detuning points.}}
        \label{fig:mixingdetune}
\end{figure}

Motivated by the above three-wave mixing result, we also qualitatively investigated the potential of the flux bias resonator to be used as a parametric amplifier. First, we operated the amplifier with non-degenerate three-wave pumping where the signal is detuned \SI{4}{kHz} above resonant frequency and the pump is set at twice the resonant frequency. The signal peak power is measured with a spectrum analyzer. Fig.~\ref{fig:mixingdetune} shows the measured signal power as a function of pump power at two DC biases corresponding to two frequency detunings. The increased DC flux bias results in both larger detuning and a boost of the three-wave mixing strength $\xi$. Increasing the pump power eventually results in \SI{10}{dB} gain on the signal for both detunings with the larger detuning showing amplification at lower pump amplitudes. This corresponds with the theoretical prediction that $\xi$ is linearly proportional to both $I_{dc}$ in the resonator as well as the pumping current amplitude $I_{p}$.

\begin{figure}
         \begin{overpic}[abs,unit=1pt,scale=.09,width=0.49\textwidth]{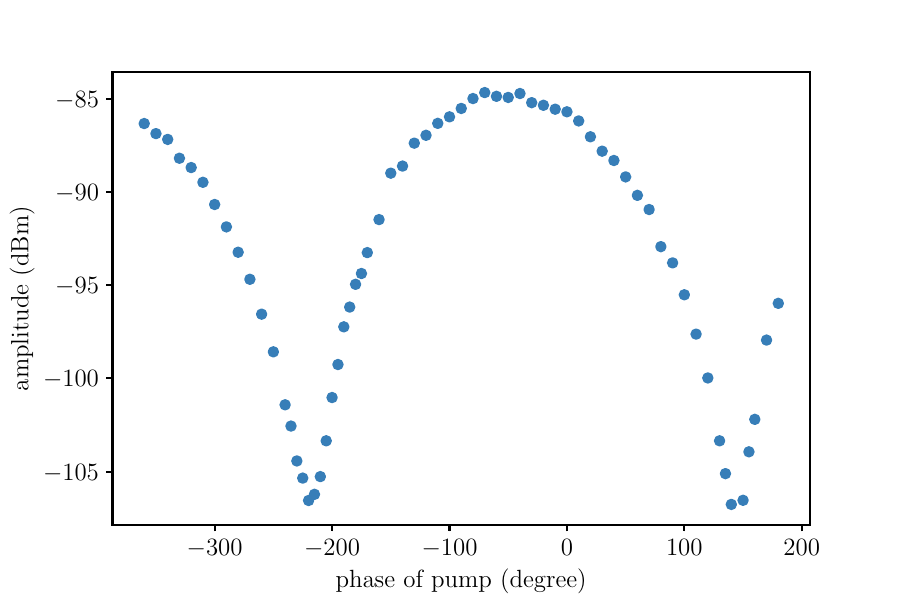}
         \end{overpic}
        \caption{\footnotesize{Measured signal peak power under degenerate three-wave pumping as function of signal phase. Both pump power and phase as well as signal output power from generator remain constant during measurement while phase of the signal was changed by adjusting the phase parameter on the generator. Power of pump was set at -51dBm.}}
        \label{fig:mixingsphase}
\end{figure}

We also characterized the device gain under degenerate three-wave mixing, where a signal tone at the resonator frequency and a pump tone at twice the resonator frequency are applied. Under this condition the pump signal splits into pairs of signal and idler photons that are at the same frequency and the system gain varies with the phase difference between the pump and signal. Results of the measured signal peak as a function of the pump tone phase are shown in Fig.~\ref{fig:mixingsphase}. Since the signal phase remains constant, changing the pump phase changes the relative phase between the pump and the signal and the measurement clearly demonstrates dependence of the amplification on the pump phase with a period of $\pi$. On the other hand, we varied the phase of the signal while keeping the pump phase constant and observed the expected change in amplification modulation period from $\pi$ to $\pi/2$.   

\vspace*{-0.3cm}
\section{Conclusion}
\label{sec::con}
In conclusion, we have demonstrated the flux-bias tuning of the resonant frequency of a superconducting resonator by using the screening current induced by magnetic flux through a superconducting loop in combination with the non-linear inductance of a superconducting thin film. Our devices exhibits the expected frequency detuning as demonstrated by measurements of the frequency response for varying DC flux bias, along with the mixing of low frequency ($\sim$MHz) and high frequency ($\sim$GHz) signals. Our measurements also provides an initial demonstration of potential applications of this architecture including upconverting MHz signals and parametric amplification through three-wave mixing. Larger frequency detuning can potentially be achieved by fabricating devices with narrower and thinner loop wires and using other superconducting films such as Titanium nitride or Nibium nitride, which have very high kinetic inductance. Devices using these alternative materials may also enable operation at temperatures above 4~Kelvin.

\section*{Data Availability Statement}
Data used in this work is available on reasonable request.
\vspace*{-0.3cm}
\section*{Acknowledgement}
We thank Professor Jarryd Pla from UNSW of Sydney, Australia for very helpful discussion on system Hamiltonian construction. Work at Argonne National Lab, including work performed at the Center for Nanoscale Materials, a U.S. Department of Energy Office of Science User Facility, is supported by the U.S. Department of Energy, Office of Science, Office of High Energy Physics and Office of Basic Energy Sciences, under Contract No. DE-AC02-06CH11357. This material is based upon work supported by the U.S. Department of Energy Office of Science National Quantum Information Science Research Centers. The work at Q-Next includes concept development, design, fabrication, testing, and modeling of devices.
\vspace*{-0.3cm}
\appendix
\section{Calculation of $df/d\Phi$}
\label{app::dfdphi}

\begin{table}[b]
\caption{\label{tab:table2}
mathmatical symbols and corresponding definition}
\begin{ruledtabular}
\begin{tabular}{ll}
 symbol & definition \\
\hline
$\omega_{o}$ & angular frequency of flux biased resonator \\
$\omega_{m}$ & angular frequency of modulation tone \\
$L_{k}$ & kinetic inductance of Al inductor  \\
$L_{k,o}$ & intrinsic kinetic inductance of Al inductor \\
$I_{sc}$ & screening current of superconducting loop \\
$I^{*}$ & characteristic current of Al inductor \\
$L_{\text{self}}$ & self inductance of superconducting loop \\
$\Phi_{dc}$ & external flux threading the supoerconducting loop\\
$\Phi_{o}$ & flux quantum \\
$I_{\text{dc}}$ & DC bias current in the readout line \\
$\overline{I}_{\text{dc}}$ & DC current in superconducting loop \\
$\overline{I}_{p}$ & pump current in superconducting loop \\
$L_{o}$ & geometric inductance of inductor \\
$L_{T}$ & total inductance of inductor \\
$\alpha$ & kinetic inductance participation ratio \\
$M$ & mutual inductance between readout line and loop\\
$K$ & Kerr nonlinearity of the flux biased resonator \\
$Q_{o}$ & internal quality factor \\
$Q_{c}$ & external of coupling quality factor 
\end{tabular}
\end{ruledtabular}
\end{table}

Complete solution for self inductance of a rectangular loop is given as~\cite{clayton2009}
\begin{equation}
    L_{\text{self}} = \frac{\mu_{o}\mu_{r}}{\pi}\left[-2(w+h)+2\sqrt{h^{2}+w^{2}}+\text{temp}\right]
\label{eq::lloop}
\end{equation}
with
\begin{eqnarray}\nonumber
    \text{temp} &=& -h~\text{ln}\left(\frac{h+\sqrt{h^{2}+w^{2}}}{w}-\right)-w ~\text{ln}\left(\frac{w+\sqrt{h^{2}+w^{2}}}{h}\right) \\
    &&+h ~\text{ln}\left(\frac{2h}{d/2}\right)+w~\text{ln}\left(\frac{2w}{d/2}\right) \nonumber
\end{eqnarray}
where $w$, $h$ are the width and height of the inductor square loop and $d$ is the diameter of the inductor wire. We will ignore the small correction to this expression due to the presence of a ground plane~\cite{jia2015}.

The resonant frequency of an LC-resonator is given by:
\begin{eqnarray}
f_{t} &=& \frac{1}{2\pi\sqrt{L_{t}C_{t}}} = \frac{1}{2\pi\sqrt{(L_{o}+L_{k,o}+\Delta L_{k})C_{t}}}\\
&=& \frac{1}{2\pi\sqrt{(L_{o}+L_{k,o})C_{t}}}\left(1-\frac{\Delta L_{k}}{2(L_{o}+L_{k,o})}\right) \\
&=& f_{o}\left(1-\frac{\Delta L_{k}}{2(L_{o}+L_{k,o})}\right)
\end{eqnarray}
with $f_{o}=1/(2\pi\sqrt{(L_{o}+L_{k,o})C_{t}})$ and $\Delta L_{k}=L_{k,o}\left(I_{sc}/I^{\ast}\right)^{2}$

The relative frequency shift due to change in kinetic inductance is expressed as:
\begin{eqnarray}
\frac{\Delta f}{f_{o}} &=& \frac{f-f_{o}}{f_{o}}=\frac{-\Delta L_{k}}{2(L_{0}+L_{k,0})} \\
&=& \frac{L_{k,0}}{2(L_{0}+L_{k,0})}\left(\frac{I_{sc}}{I^{\ast}}\right)^{2} \\
&=& \frac{L_{k,o}}{2(L_{o}+L_{k,o})}\left(\frac{I_{sc}}{I^{\ast}}\right)^{2}
\label{eq:frefractionIsc}
\end{eqnarray}
Expressing the screening current $I_{sc}$ from Eq.~\ref{eq::lloopphi}
\begin{eqnarray}
I_{sc} &=& \frac{-\Phi_{dc}+m\Phi_{o}}{L_{\text{self}}} \\
&=& \frac{-M I_{dc}+m\Phi_{o}}{L_{\text{self}}}
\label{eq::Iscphi}
\end{eqnarray}
where $M$ is the mutual inductance between the readout line and the superconducting loop. substituting $I_{sc}$ into Eq.~\ref{eq::Iscphi} we obtain:
\begin{eqnarray}
\frac{\Delta f}{f_{o}} &=& \frac{L_{k,o}}{2L_{\text{self}}}\left(\frac{I_{sc}}{I^{\ast}}\right)^{2} \\
&=& \frac{L_{k,o}(-M I_{dc}+m\Phi_{o})^{2}}{2(L_{o}+L_{k,o})L_{\text{self}}^{2}I^{\ast 2}} \\
&=& \frac{\alpha(-M I_{dc}+m\Phi_{o})^{2}}{2L_{\text{self}}^{2}I^{\ast 2}}
\label{eq:sss}
\end{eqnarray}
The derivative of frequency change versa current bias is:
\begin{eqnarray}
\frac{d f}{dI_{dc}} &=& \frac{f_{o}\alpha M^{2}}{L_{\text{self}}^{2}I^{\ast 2}}I_{dc}
\label{eq::dfdt}
\end{eqnarray}
where we have set the intrinsic loop flux $m=0$.
\vspace*{-0.3cm}
\section{optimization of $df/d \Phi$}
several geometrical factors play role in determining the relative magnitude of resonance frequency shift induced by the $I_{DC}$: the resonator geometry such as the width and length of the loop and the cross-section of the superconducting microstrip, as well as the mutual inductance between the resonator loop and the DC line. 
 In our resonator design $Q_{c}\ll Q_{o}$ so the total quality factor $Q_{r}$ is dominated by $Q_{c}$. We choose the width of the loop to be \SI{500}{\mu m} because it gives us the high $Q_{c}$ based on our related resonator design. It will be easier to achieve flux bias with a wider loop but at the same time $Q_{c}$ will decrease. The gap between the resonator and the readout line is \SI{2}{\mu m}. A smaller gap will have more flux bias but it would also bring challenges related to lithographic patterning of the structure. At this point we prefer to limit ourself with the best resolution of our optical lithography. Thus, in our designs we fixed the width to be \SI{500}{\mu m} and gap to be \SI{2}{\mu m}.

\begin{figure}[h!]
     \begin{overpic}[abs,unit=1pt, scale=.35,width=0.51\textwidth]{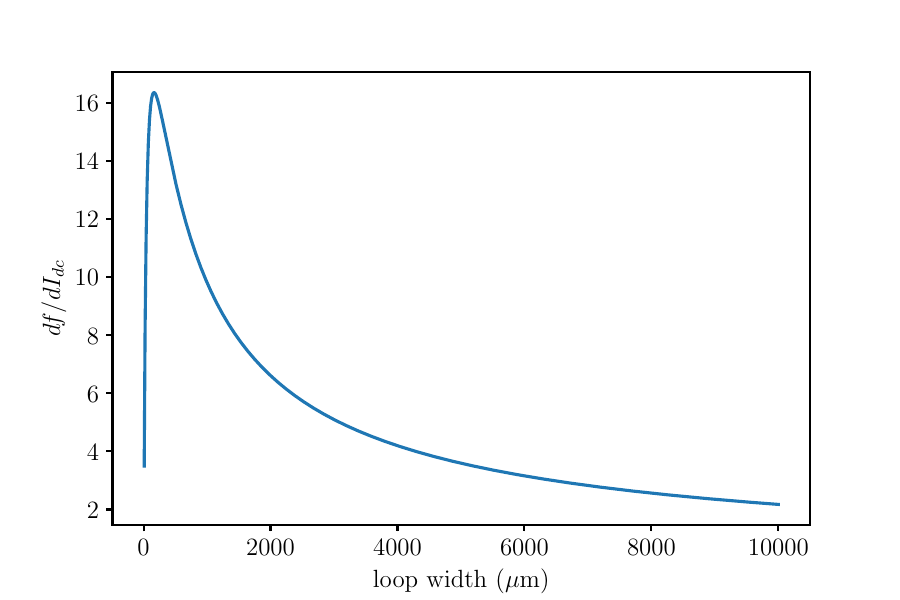}
     \put(85, 60){\color{white}
     \frame{\includegraphics[scale=0.55]{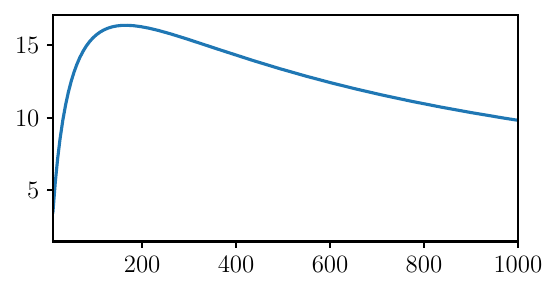}}}
     \end{overpic}\
     \vskip 3pt
     \caption{\footnotesize{$\partial f/\partial I_{dc}$ as function of loop width W for loop strip of \SI{1}{\mu m}. Maximum happens at W~$\approx$~\SI{200}{\mu m}. Inset is zoom in to the range of \SI{1000}{um} to show the maximum point.}}
\label{fig::optimal}
\end{figure}

In Fig.~\ref{fig::optimal} we plot the $df/d I_{dc}$ as function of the loop height $H$ from 0 to \SI{10000}{\mu m} for loop microstrip diameter of \SI{1}{\mu m} and the maximum frequency shift happens at $H\approx$~ \SI{200}{\mu m}. This gives a guideline regarding the target parameters of the resonator design.
\vspace*{-0.3cm}
\section{nonlinear resonator model}
\label{ap::resoham}
For kinetic inductance based nonlinear resonator
\begin{eqnarray}
L_{k}(I) &=& L_{k, o}\left(1+\frac{I^{2}}{I^{\ast 2}}\right)
\end{eqnarray}
calculated energy of the resonator is as follows:
\begin{eqnarray}
U_{C} &=& \frac{1}{2}C\dot{\Phi}^{2} \\
U_{L} &=& \frac{1}{2}\frac{\Phi^{2}}{L_{k}+L_{o}}\\
&=& \frac{1}{2}\frac{\Phi^{2}}{L_{o,k} \left(1+\frac{I^{2}}{I^{\ast 2}}\right)+L_{o}}\\
&=& \frac{1}{2}\frac{\Phi^{2}}{L_{T}}-\frac{L_{o,k}}{2L_{T}^{4}}\frac{\Phi^{4}}{I^{\ast 2}}
\end{eqnarray}
where $L_{T} = L_{o} + L_{k,o}$. $L_{o}$ and $L_{k, o}$ is the geometric and kinetic inductance without current bias respectively. Lagrangian of the resonator is given as
\begin{eqnarray}
\mathcal{L} &=& U_{C} - U_{L} = \frac{1}{2}C\dot{\Phi}^{2}-\frac{1}{2}\frac{\Phi^{2}}{L_{T}}+\frac{L_{o,k}}{2L_{T}^{4}}\frac{\Phi^{4}}{I^{\ast 2}}
\end{eqnarray}
Using Legendre transformation $\mathcal{H} = \dot{\Phi}Q - \mathcal{L}$ the Hamiltonian is written as follows
\begin{eqnarray}
\mathcal{H} &=& \frac{Q^{2}}{2C} + \frac{1}{2}\frac{\Phi^{2}}{L_{T}}-\frac{L_{o,k}}{2L_{T}^{4}}\frac{\Phi^{4}}{I^{\ast 2}}
\end{eqnarray}
Replacing $Q$ and $\Phi$ with their quantum operator the Hamiltonian now can be written as
\begin{eqnarray}
\mathcal{\tilde{H}} &=& \frac{\tilde{Q}^{2}}{2C} + \frac{1}{2}\frac{\tilde{\Phi}^{2}}{L_{T}}-\frac{L_{o,k}}{2L_{T}^{4}}\frac{\tilde{\Phi}^{4}}{I^{\ast 2}}
\end{eqnarray}
we can further separate the Hamiltonian into its linear $\mathcal{\tilde{H}}_{o}$ and interaction $\mathcal{\tilde{H}}_{I}$ part
\begin{eqnarray}
\mathcal{\tilde{H}}_{o} &=& \frac{\tilde{Q}^{2}}{2C} + \frac{1}{2}\frac{\tilde{\Phi}^{2}}{L_{T}} \\
\mathcal{\tilde{H}}_{I} &=& -\frac{L_{o,k}}{2L_{T}^{4}}\frac{\Phi^{4}}{I^{\ast 2}}
\end{eqnarray}

Defining the following creation and annihilation operators with $\omega_{o} = 1/\sqrt{L_{T}C}$
\begin{eqnarray}
\tilde{a} &=& \frac{1}{\sqrt{\hbar\omega_{o}}}\left [\frac{1}{\sqrt{2L_{T}}}\tilde{\Phi}+i\frac{1}{\sqrt{2C}}\tilde{Q}\right ] \\
\tilde{a}^{\dagger} &=& \frac{1}{\sqrt{\hbar\omega_{o}}}\left [\frac{1}{\sqrt{2L_{T}}}\tilde{\Phi}-i\frac{1}{\sqrt{2C}}\tilde{Q}\right ]
\end{eqnarray}
the reduced flux operator in terms of $\tilde{a}$ and $\tilde{a}^{\dagger}$
\begin{eqnarray}
\tilde{\Phi} &=& \sqrt{\hbar\omega_{o} L_{T}/2}(\tilde{a}^{\dagger}+ \tilde{a})
\end{eqnarray}
and the interaction Hamiltonian is
\begin{eqnarray}
\mathcal{H}_{I} &=& -\frac{L_{o,k}}{2L_{T}^{4}}\frac{1}{I^{\ast 2}} (\hbar\omega_{o} L_{T}/2)^{2}(\tilde{a}^{\dagger}+ \tilde{a})^{4} \\
&=& -\frac{3L_{o,k}(\hbar\omega_{o} L_{T})^{2}}{4L_{T}^{4}I_{\ast}^{2}} (2\tilde{a}^{\dagger}\tilde{a} + \tilde{a}^{\dagger 2} \tilde{a}^{2}) \\
&=& \hbar K \tilde{a}^{\dagger}\tilde{a} + \frac{1}{2}\hbar K \tilde{a}^{\dagger 2} \tilde{a}^{2}
\end{eqnarray}
 and $K = -\frac{3\hbar\omega_{o}^{2} L_{o,k}}{2L_{T}^{2}I_{\ast}^{2}} = -\frac{3\hbar\omega_{o}^{2}\alpha}{2L_{T}I_{\ast}^{2}}$. $\alpha$ is the participation  ratio of the kinetic inductance \\
\\
with DC bias  $\overline{I}_{dc}$ and parametric pump mode $\overline{I}_{p}$ in the resonator
\begin{widetext}
\begin{eqnarray}
\mathcal{H}_{I} &=& \frac{1}{2}\hbar K\left(2a^{\dagger}a+a^{\dagger^{2}}a^{2}\right)- \frac{L_{o,k}}{2L_{T}^{4}}\frac{1}{I^{\ast 2}}6(\hbar\omega_{o} L_{T}/2)L_{T}^{2}\left[\overline{I}_{dc}^{2}+2\overline{I}_{dc}\overline{I}_{p}(t)+\overline{I}_{p}^{2}(t)\right]\left(2a^{\dagger}a+a^{
\dagger^{2}}+a^{2}\right) \\
&=& \frac{1}{2}\hbar K\left(2a^{\dagger}a+a^{\dagger^{2}}a^{2}\right) - \frac{3\hbar\omega_{o}\alpha}{2I_{\ast}^{2}}\left[\overline{I}_{dc}^{2}+2\overline{I}_{dc}\overline{I}_{p}(t)+\overline{I}_{p}^{2}(t)\right]\left(2a^{\dagger}a+a^{
\dagger^{2}}+a^{2}\right)
\end{eqnarray}
\end{widetext}
the time dependent pump mode can be written as
\begin{eqnarray}
\overline{I}_{p}(t) &=& \frac{\overline{I}_{p}}{2}\left(e^{-i(\omega_{p}t+\phi_{p})}+e^{i(\omega_{p}t+\phi_{p})}\right)
\end{eqnarray}
Apply rotating wave approximation (RWA)
\begin{eqnarray}
\mathcal{H}_I' &=& U^\dagger \mathcal{H}_I U - i\hbar U^\dagger \frac{\partial}{\partial t}U 
\end{eqnarray}
and 
\begin{eqnarray}
U^\dagger &=& \exp(i\omega_p t a^\dagger a/2) \\
U &=& \exp(-i\omega_p t a^\dagger a/2)
\end{eqnarray}
We transfer the Hamiltonian into frame rotating at $\omega_{p}/2$ 
\begin{eqnarray}
\mathcal{H} &=& \hbar (\omega_{o}+ \delta_{dc}+\delta_{p}+ K - \frac{\omega_{p}}{2})a^{\dagger} a \\
&& + \frac{\hbar \xi}{2}a^{\dagger 2} + \frac{\hbar \xi^{\ast}}{2}a^{2} + \frac{\hbar K}{2} a^{\dagger 2} a^{2}
\end{eqnarray}
\begin{eqnarray}
\delta_{dc} &=& -\frac{3\omega_{o}\alpha \overline{I}_{dc}^{2}}{I_{\ast}^{2}} \\
\delta_{p} &=& -\frac{3\omega_{o}\alpha \overline{I}_{p}^{2}}{4I_{\ast}^{2}} \\
K &=& -\frac{3\hbar\omega_{o}^{2}\alpha}{2L_{T}I_{\ast}^{2}} \\
\xi &=& -\frac{3\omega_{o}\alpha \overline{I}_{dc}\overline{I}_{p}}{2I_{\ast}^{2}}e^{-i\phi_{p}}
\end{eqnarray}

the system Hamiltonian can be calculated by setting the pump amplitude to zero and 
\begin{eqnarray}
\label{eq::hamdfdi}
\mathcal{H} &=& \hbar (\omega_{o}+ \delta_{dc}+ K )a^{\dagger} a + \frac{\hbar K}{2} a^{\dagger 2} a^{2} \\
\label{eq::hamIdc}
\delta_{dc} &=& -\frac{3\omega_{o}\alpha \overline{I}_{dc}^{2}}{I_{\ast}^{2}}
\end{eqnarray}
\bibliography{reff}
\end{document}